# Tests on the POD-P controller of INELFE Spain-France VSC-HVDC interconnector

Javier Renedo
Red Eléctrica, Spain
javier.renedo@ree.es

Agustín Díaz-García
Red Eléctrica, Spain
agustin.diaz@ree.es

Gilles TORRESAN
RTE, France
gilles.torresan@rte-france.com

Eduardo Lorenzo Cabrera
Red Eléctrica, Spain
elorenzo@redeia.com

Antonio Cordón
Red Eléctrica, Spain
acordon@ree.es

Silvia Sanz Verdugo
Red Eléctrica, Spain
sisanz@ree.es

Juan Peiró
Red Eléctrica, Spain
jpeiro@ree.es

Javier Pérez Castro
Red Eléctrica, Spain
japerez@ree.es

Gorka Calvo
Red Eléctrica, Spain
gorka.calvo@ree.es

Aitor Hernández Sautua
Red Eléctrica, Spain
aitor.hernandez@ree.es

Ambroise PETIT
RTE, France
ambroise.petit@rte-france.com

Hamza OUKHAYI
RTE, France
hamza.oukhayi@rte-france.com

Samy AKKARI
RTE, France
samy.akkari@rte-france.com

Hani SAAD
France

## SUMMARY

INELFE interconnector consists of a 2x1000 MW high voltage direct current system based on voltage source converters (VSC-HVDC) interconnecting France and Spain. INELFE VSC-HVDC link is embedded into the high voltage alternating current (HVAC) system. Electromechanical oscillations, also known as power oscillations, are a major concern worldwide. INELFE VSC-HVDC link has specific controllers to damp power oscillations by modulating active (P)- and reactive (Q)-power injections of the VSC converters (POD-P and POD-Q controllers, respectively).
The Spanish and French Transmission System Operators (TSOs) carried out a join task force to

- Increase the gain of POD-P controller of INELFE VSC-HVDC link.
- Make it possible to use the POD-P controller together with angle difference control (ADC) of INELFE VSC-HVDC interconnector.

The objective of these modifications is to improve the effectiveness of POD-P controller and, therefore, to increase its contribution to the damping of inter-area oscillations in the Continental Europe (CE) power system. Such changes require (a) extensive simulation studies and (b) extensive tests in different operation modes, in order to ensure the correct behavior of the system.
This paper presents simulation studies and field tests on the POD-P controller of INELFE VSC-HVDC interconnector in different modes of operation.

# 1 INTRODUCTION,

Continental Europe (CE) power system is one of the largest synchronous power system in the world. It supplies energy to the continental European Countries from Spain and Portugal (Iberian Peninsula) in the Southwest to Denmark, Poland in the North and Romania and Bulgaria in the East. Turkey and Ukraine were connected to the CE power system in 2010 and during 2022, respectively.

In such a large power system, inter-area oscillations are a major concern. There are three main inter-area modes in CE: North-South, East-West and East-Center-West. Thus, the whole of the system is involved in this phenomenon. Some oscillatory events in the last years involving inter-area oscillations ( [1], [2], [3]) have evidenced the need to address properly this challenge and mobilize all the resources available in the system to help to damp inter-area oscillations.

INELFE interconnector consists of a 2x1000 MW high voltage direct current system based on voltage source converters (VSC-HVDC) interconnecting France and Spain through the East Pyrenees. It was put in service in 2015, becoming the Europe's first HVDC underground interconnector with VSC technology integrated into a meshed AC grid ( [4], [5]). Transmission system operators (TSO) of France and Spain (Réseau de Transport d'Électricité (RTE) and Red Eléctrica (RE), respectively) have worked together since the commissioning of INELFE HVDC interconnector, in order to try to improve its active and reactive power controllers to have a positive impact on the damping ratio of inter-area oscillations.

VSC-HVDC systems can help to damp electromechanical oscillations by means of power oscillation damping (POD) controllers for the active and reactive-power injections of the converter stations (POD-P and POD-Q controllers, respectively) [6], [7], [8], [9]. Meanwhile, HVDC systems based on Line Commutated Converters (LCC-HVDC) also proved to be effective to damp inter-area oscillations using POD-P controllers [10].

INELFE VSC-HVDC interconnector is equipped with POD-P and POD-Q controllers. Due to the location of the HVDC system and, mainly, because the HVDC is almost in parallel with an AC corridor, the effectiveness of the POD-P controller is limited. Nevertheless, in scenarios in which this parallel AC corridor is open (e.g. due to maintenance), the effectiveness of POD-P controller increases significantly. When the parallel AC corridor is open, the electrical distance from the Iberian Peninsula to the center of the power system increases and, therefore, the damping of the inter-area modes tends to be lower, in general. Hence, having available POD-P controller in this situation and in an effective way is of great interest.

In normal operation, INELFE HVDC interconnector is operated with Angle Difference Control (ADC) [4], [5], also known as AC-line emulation control, because it allows to operate easily a VSC-HVDC system embedded into an AC system. Previous work has shown that ADC control in VSC-HVDC links could have negative impact on the damping ratio of inter-area oscillations ( [11], [12], [13], [14]). To improve the oscillatory behavior of the system, ADC control of INELFE HVDC interconnector was made slower by increasing its time constant from $\tau = 0.75$ s to $\tau = 50$ s [11], [12].

Originally, the activation of POD-P controller together with ADC mode was forbidden by the HVDC control, in order to avoid interactions between both controllers. After the slowdown of ADC, the risk of interaction is decreased significantly.

Then, the Spanish and French TSOs carried out a join task force to:

- Increase the gain of POD-P controller, in order to improve its effectiveness.
- Allow the activation of POD-P controller together with angle difference control (ADC).

The objective of these modifications is to improve the effectiveness of POD-P controller and, therefore, to increase its contribution to the damping of inter-area oscillations in the CE power system. These changes require (a) extensive simulation studies and (b) extensive tests in different operation modes, in order to ensure the correct behavior of the system.
This paper presents simulation studies, implementation aspects and field tests on the POD-P controller of INELFE VSC-HVDC interconnector. Results prove the correct behavior of the POD-P controller in different modes of operation.

## 2 DESCRIPTION OF THE SYSTEM

Figure 1 shows the transmission network in the French-Spanish border and the AC/DC corridor formed by the HVDC interconnector and line Vic-Baixas 400 kV. The interconnection capacity between the Iberian Peninsula (Spain and Portugal) and the rest of the CE system is 2.800 MW, provided by four AC transmission lines between Spain and France (two 400 kV lines and two 220 kV lines) together with an HVDC interconnector through the Eastern Pyrenees (INELFE interconnector).

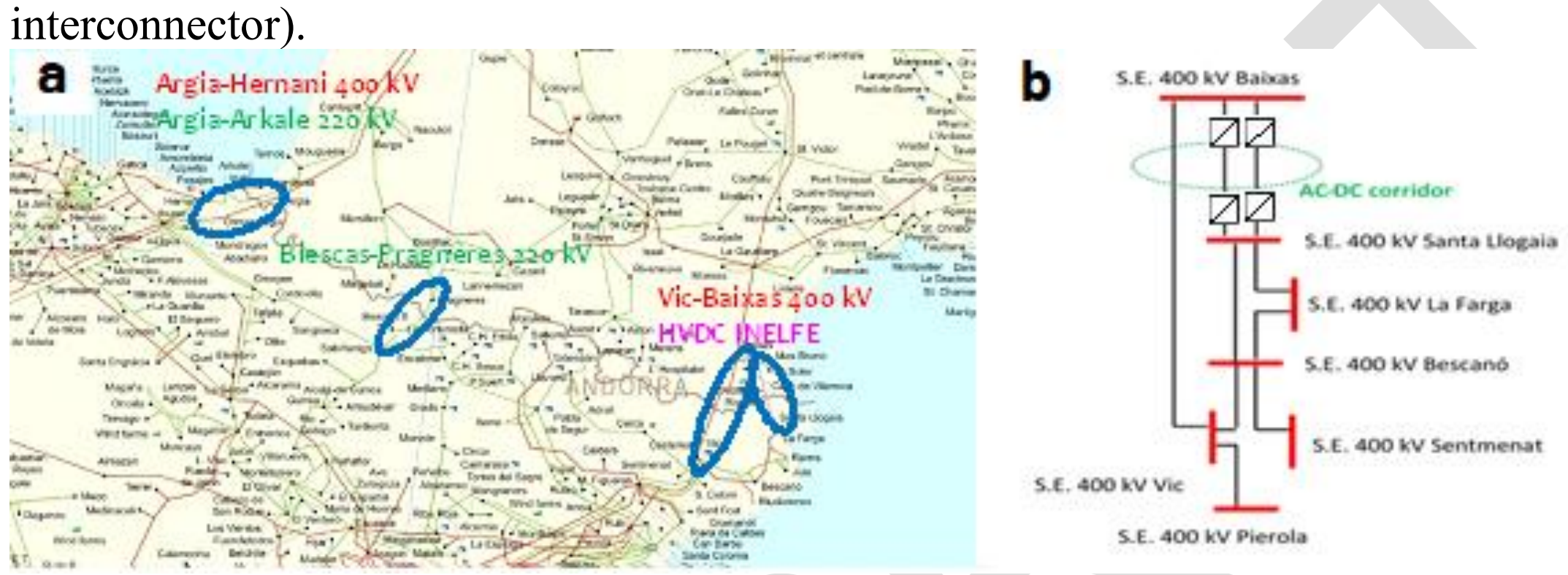

**Figure 1 : (a) Transmission network in the French-Spanish border and (b) AC/DC corridor formed by the HVDC interconnector and line Vic-Baixas 400 kV.**

INELFE VSC-HVDC interconnector is composed of two 1000 MW symmetrical monopoles at 320 kV DC. Each converter has a reactive power capability at the connection point of [-600 Mvar, +400 Mvar]. The peculiarity of being a HVDC system embedded in the CE system, practically in parallel with line Vic-Baixas 400 kV and forming one of the most important corridors for active power transmission between France and Spain, requires a special operation. Thus, the active power flow through the line Vic-Baixas 400 kV is highly sensitive to changes in the HVDC active power setpoint: around 70% of the change made in the HVDC active power setpoint is assumed by the line Vic-Baixas 400 kV, the remaining 30% of the change being distributed among the other three AC interconnection lines.
The VSC-HVDC system is highly controllable in active power flow and reactive power injections. Figure 2 shows a general scheme of INELFE HVDC interconnector.

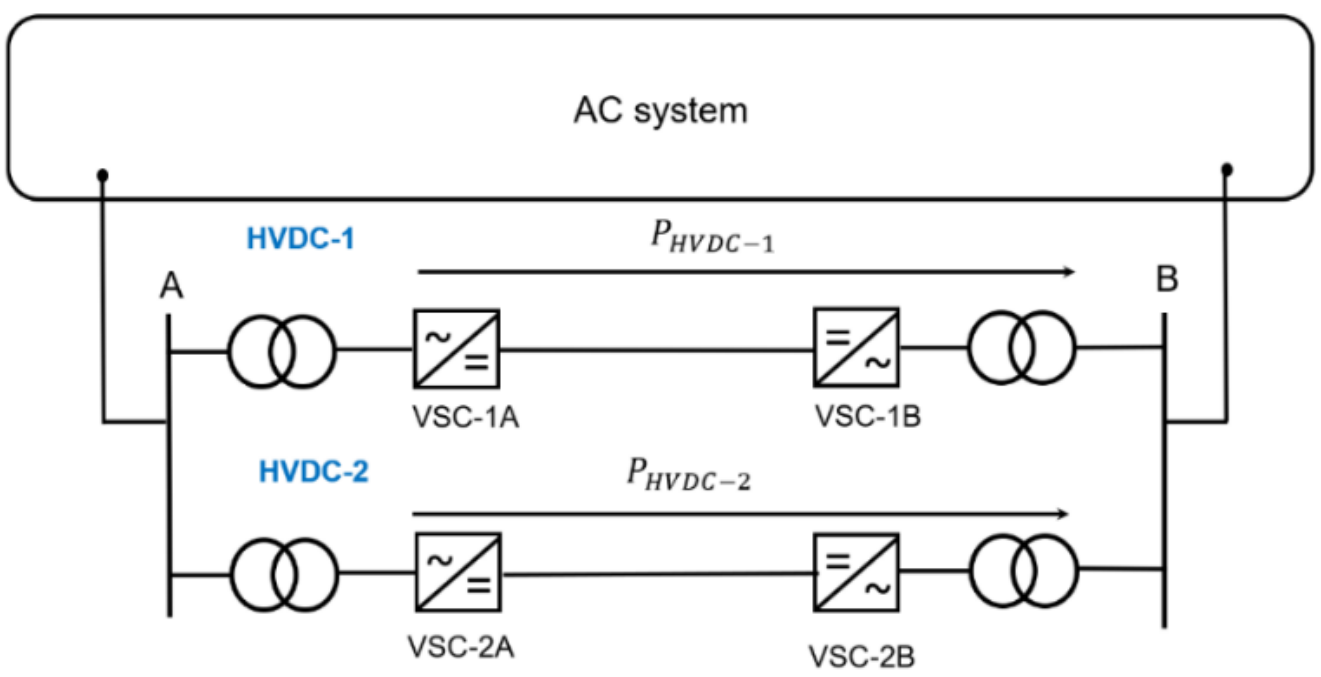

**Figure 2 : General scheme of INELFE HVDC interconnector. Bus A : Santa LLogaia 400 kV (Spain). Bus B : Baixas 400 kV (France).**

There are two controls of the active-power flow of the HVDC system:

- Constant power control (CPC), also known as Pmode1 control.
- Angle difference control (ADC), also known as Pmode3 control.

## 2.1 Constant power control (CPC)

In CPC, each HVDC link controls its active power flow to a constant value. Hence, the total active power set point of each HVDC link is given by:

$$P^*_{HVDC-i} = P_0 + \Delta P^{ref,POD} \quad \text{(1)}$$

where:

- $P_0$ is a constant active-power set point.
- $\Delta P^{ref,POD}$ is the supplementary active-power set point provided by POD-P controller.

## 2.2 Angle difference control (ADC)

In ADC, each HVDC link controls its active power flow with a term proportional to the angle difference between the two AC terminals of the HVDC link, emulating the behavior of an AC line. When an HVDC is embedded in an AC network, its active power transfer has a relevant influence on the rest of the AC network. In these situations, ADC control is more practical than CPC control. In ADC, the total active power set point of each HVDC link is given by:

$$P^*_{HVDC-i} = P_0 + \Delta P^{ref,ADC} + \Delta P^{ref,POD} \quad \text{(2)}$$

with:

$$\Delta P^{ref,ADC} = \frac{K}{1 + s\tau}(\delta_A - \delta_B) \quad \text{(3)}$$

where:

- $P_0$ is a constant active-power set point.
- $\Delta P^{ref,ADC}$ is the supplementary active-power set point provided by ADC controller.
- $\delta_A$ and $\delta_B$ are the angles of the voltages at Santa Llogaia 400 kV (Spain) and at Baixas 400 kV (France), respectively.
- $\Delta P^{ref,POD}$ is the supplementary active-power set point provided by POD-P controller.

**Table 1 : ADC control: parameters (per link).**

| Parameter | Description | Value |
|---|---|---|
| $P_0$ | Constant active-power set point. | 0 MW |
| $K$ | Gain of ADC. | 180 MW/deg |
| $\tau$ | Time constant of ADC | 50 s |

## 2.3 POD-P control

Figure 3 shows the block diagram of POD-P controller. The input signal of POD-P controller is the difference of the frequencies between the AC terminals of the VSC-HVDC system: $\omega_A$ and $\omega_B$ are the frequencies at Santa Llogaia 400 kV (Spain) and at Baixas 400 kV (France), respectively. POD-P controller contains a gain, a low-pass filter, a wash-out filter and a saturation parameter, as summarised in Table 2. The output of the controller is a supplementary active-power setpoint for each HVDC link, $\Delta P^{ref,POD}$.

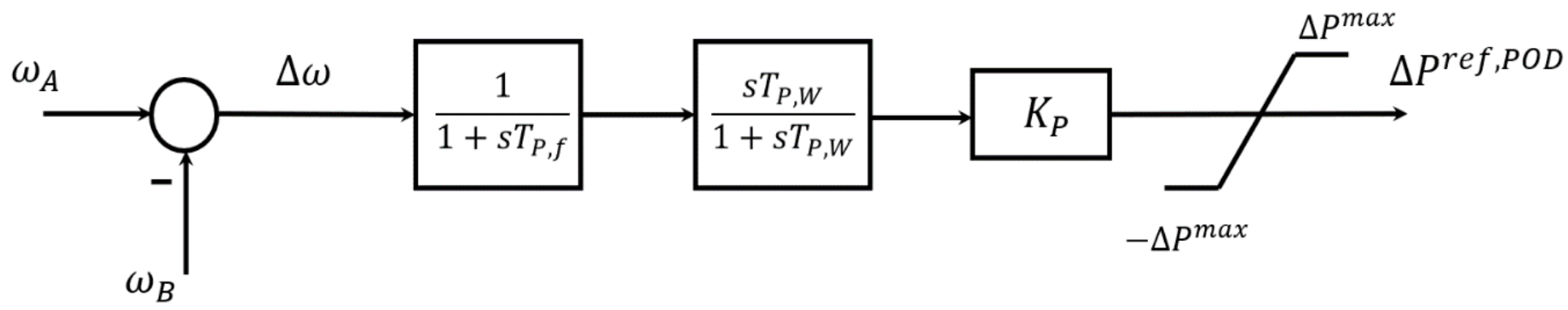

**Figure 3 : POD-P controller.**

**Table 2 : POD-P control: parameters (per link).**

| Parameter | Description | Value |
|---|---|---|
| $K_P$ | Controller gain. | Original: 500 MW/Hz<br>New: 2500 MW/Hz |
| $T_{P,f}$ | Time constant of the low-pass filter. | 0.15 s |
| $T_{P,W}$ | Time constant of the wash-out filter. | 2.5 s |
| $\pm\Delta P^{max}$ | Saturation parameter. | $\pm 100$ MW |

Originally,

- Gain of POD-P controller was set to: $K_P = 500$ MW/Hz (per link).
- POD-P controller was automatically disabled when operating in ADC mode ($\Delta P^{ref,POD} = 0$ MW, always). The purpose of this logic condition was to avoid interactions between ADC and POD-P controller.

The following modifications were made, in order to improve the effectiveness of POD-P controller:

- Gain of POD-P controller was increased to: $K_P = 2500$ MW/Hz (per link).
- Modification in the logic conditions in order to be able to activate POD-P controller in ADC mode. Since ADC controller was made slower ($\tau = 50$ s) [12], ADC and POD-P controllers are decoupled and there is no risk of interaction.

The changes made on POD-P controller were supported by power system studies, hardware-in-the-loop studies (in replica) and field tests.

## 3 POWER SYSTEM STUDIES ON POD-P CONTROLLER

For the dynamic studies, Red Eléctrica used the Dynamic Study Model – DSM- [15], while RTE used Dynamic Reference Model – DRM- [16]. Both models, after appropriate tuning, were able to reproduce accurately the East-Center-West inter-area oscillation in terms of frequency, damping ratio and mode shape, thus, allowing to perform the study of the impact of INELFE HVDC control on this mode. Red Eléctrica uses Powerfactory (DIgSILENT) [17] for small-signal stability studies and PSS/E [18] for transient stability studies. RTE uses SMAS3 [19] for small signal stability studies and Eurostag [20] for transient stability studies. For the sake of clarity, only the results obtained by RTE are presented in this paper.

**Eigenvalue analysis**

Table 3 shows results based on a situation where France is importing 2000 MW from Spain in CPC mode, obtained by means of small-signal analysis using SMAS3. Results show that POD-P controller increases the damping ratio of the inter-area mode, and it is much more effective when line Vic-Baixas 400 kV is open.

**Table 3 : Eigenvalue analysis: Inter-area mode.**

| East-Center-West mode | Base case | Line Hernani-Argia 400 kV open | Line Vic-Baixas 400 kV open |
|---|---|---|---|
| POD-P OFF | Frequency: 0.23 Hz<br>Damping: 2.4% | Frequency: 0.22 Hz<br>Damping: 2.4% | Frequency: 0.20 Hz<br>Damping: 5.9% |
| POD-P ON (gain of 2500 MW/Hz/link) | Frequency: 0.23 Hz<br>Damping: 2.9% | Frequency: 0.22 Hz<br>Damping: 2.9% | Frequency: 0.20 Hz<br>Damping: 15% |

**Time domain simulations**

Extensive simulations were carried out to verify the behavior of the system after implementing the changes on POD-P controller: full network, short circuits, outages, different levels of exchanges between France and Spain, as it is a well-known variable that impacts the damping, HVDC in ADC and CDC. The scope of these simulations was to verify the results obtained from the Eigenvalue analysis and to ensure that no adverse effects were created.

Figure 4 shows simulation results for the case where France imports 3.5 GW from Spain, HVDC in ADC control, obtained with Eurostag. Two events were simulated: Opening of Line Vic-Baixas 400 kV and opening of line Argia-Hernani 400 kV. Time-domain simulations confirm the effectiveness of POD-P controller, because it contributes to damp the inter-area oscillation.

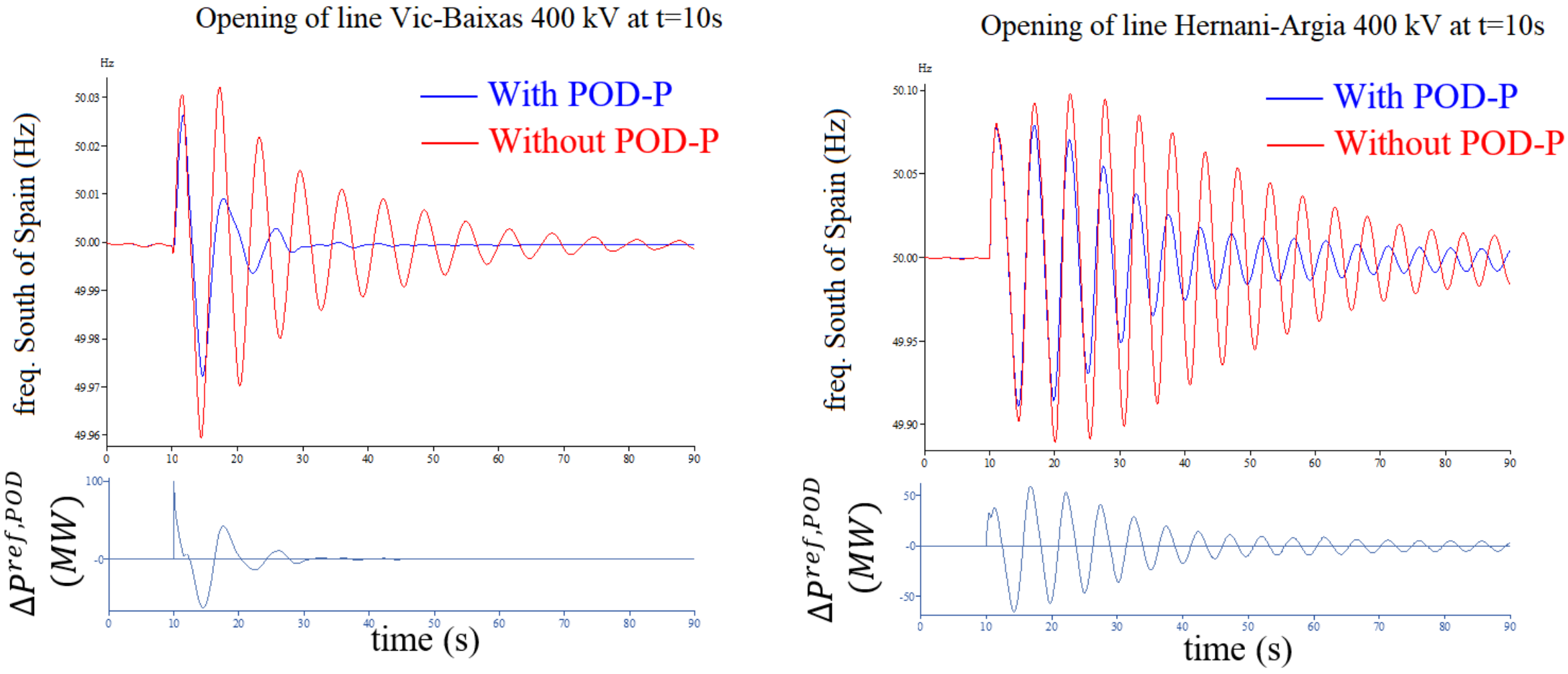

**Figure 4 : Opening of line Vic-Baixas 400 kV: (a) frequency at the South of Spain and (b) supplementary active-power set point provided by POD-P controller (gain of 2500 MW/Hz/link).**

## Discussion

Based on the frequency amplitudes experienced for historical system events, a gain of 2500 MW/Hz/link for POD-P controller seems a good compromise as:

- The system is stable.
- It would lead to saturation of POD-P at the beginning of severe events.
- It needs $\Delta\omega = \omega_A - \omega_B = 40$ mHz to saturate, which could be encountered only for very severe events.
- The damping ratio of the inter-area mode could be increased significantly when line Vic-Baixas 400 kV is opened.
- The damping ratio of the inter-area mode could be increased slightly when line Vic-Baixas 400 kV is closed.

In the simulations carried out by RTE and Red Eléctrica, gains of POD-P controller within the range $K_P = 2500 - 5000$ MW/Hz/link produced good results. However, in order to consider the inaccuracies of the dynamic models, Red Eléctrica and RTE decided to keep stability margins by using the conservative value of $K_P = 2500$MW/Hz/link. This value is consistent with typical values for POD-P controllers using frequency-error input signals [7], [9], [10].
In conclusion, for all the simulations performed the POD-P increases the damping ratio of the inter-area mode. Then, the decision was taken to implement the changes in field.

# 4 PRACTICAL IMPLEMENTATION OF THE CHANGES MADE ON POD-P CONTROLLER

To validate the feasibility of the modifications made in the POD-P controller of INELFE HVDC interconnector, their implementation in the Control and Protection (C&P) cubicles was tested tested before onsite control update. To achieve this, a hardware-in-the-loop (HIL) set-up in RTE facilities where INELFE physical replicas of C&P systems are connected to a real time simulator is utilized. Such an installation guarantees the accuracy of the HVDC C&P and has been validated several years ago [21].
A simple benchmark system (Figure 5) has been developed on the real-time simulator Hypersim (OPAL-RT) [22], [23], to reproduce the incident of 1st December 2016 [2] leading to a 0.155 Hz inter-area oscillation. Two synchronous machines represent the French (H1) and Spanish (H2) sides. The VSC-HVDC system is controlled by the INELFE replicas in the laboratory of RTE.

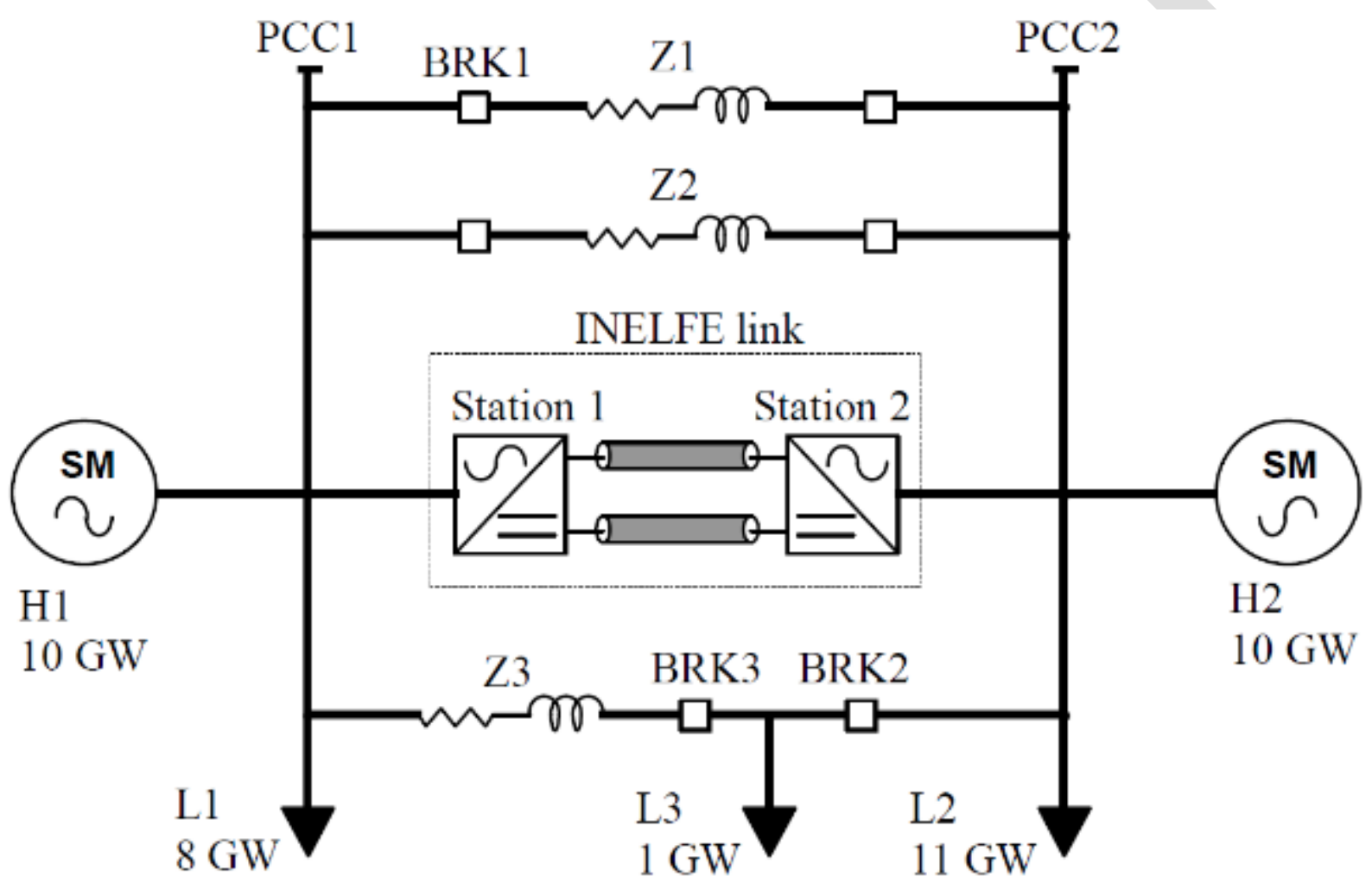

**Figure 5 : Benchmark system in Hypersim (OPAL-RT).**

Various situations have been tested: level and control mode of active or AC voltage/reactive power, inverter or rectifier mode, 0 MW crossing, value of the gain of POD-P controller. Simulations showed that the activation of the POD-P with ADC with $\tau = 50$ s was possible, and that the activation of the POD-P with a gain of $K_P = 2500$ MW/Hz/link with constant power control or ADC increases the damping ratio of the inter-area mode. Higher values of the POD-P gain were also analysed, concluding that values around $K_P = 2500 - 5000$ MW/Hz/link produced good results are they were reasonable designs. These modifications do not provoke any problems or warnings from the INELFE control and protections. Next step carried out consisted in the redaction of a joint procedure that describes how to change the POD-P control onsite by the operators (POD-P activation in ADC and change of gain to $K_P = 2500$ MW/Hz/link). The changes on POD-P controller were implemented in parallel by Red Eléctrica and RTE in September 2022, during the INELFE HVDC interconnector annual maintenance.

## 5 TESTS ON POD-P CONTROLLER

Once the changes on POD-P controller if INELFE VSC-HVDC link were implemented, extensive tests were carried out during the following months (September-December 2022), in order to check its correct behavior. The tests included different operating conditions:

- Single-link operation / Multi-link operation (e.g. the HVDC system is operated with set points for the total HVDC system, and each HVDC link is operated symmetrically in an automatic way).
- Constant active power control (CPC) / Angle difference control (ADC).
- PCONMASTER: Is the converter station that controls the active-power injection (the remaining converter station controls the DC voltage).
- Power flow direction: from Spain to France / from France to Spain.
- Disconnection of line Vic-Baixas 400 kV.

Only one representative tests are presented in this paper:

- Test 1: Tests of POD-P control in CPC and ADC (5-Oct-2022)
- Test 2: Tests of POD-P control of INELFE-1 under the disconnection of line Vic-Baixas 400 kV (9-Dec-2022).

Data obtained from Phasor Measurement Units (PMU) in Spain and in France were exported to analyze the results.

### Test 1: Tests of POD-P control in CPC and ADC (5-Oct-2022)

These tests consist in analyzing the behavior of POD-P in CPC and ADC operating modes. The objectives of this test are to evaluate the performance of POD-P against small disturbances in the system (changes on system conditions during a certain time) and to verify its correct behavior. These tests were carried out during a planned outage of line Vic-Baixas 400 kV, which is a good scenario to test the performance of POD-P controller, because, as explained in previous section, POD-P controller is much more effective when line Vic-Baixas 400 kV is open.

Two tests were carried out:

- Test 1-a: CPC and POD-P controller.
- Test 1-b: ADC and POD-P controller.

The operating conditions of the system before the tests were as follows:

- VSC-HVDC interconnection: HVDC-1 and HVDC-2 in service. Control mode: Multi-Link. France is PCONMASTER and active-power flow from Spain to France
- Line Vic-Baixas 400 kV disconnected (planned outage)

**Test 1-a: CPC with POD-P**

Figure 6 shows the difference between the frequencies of Llogaia and Baixas substations and the active-power through both HVDC links of INELFE-1 (HVDC-1 and HVDC-2) from Spain to France). The difference of the frequencies of both AC terminals is not neglectable, due to the fact that Line Vic-Baixas 400 kV is open. During the transient, when the frequency of Llogaia is greater (lower) than the frequency of Baixas, the active power through the HVDC links (from Spain to France) increases (decreases). The HVDC links maintain the average value of the active power flow constant, because they control their power flow to a constant set-point value (CPC) plus the active-power set point of the POD-P controller. Hence, the POD-P controller presents a correct behavior. Notice that the behavior of the active power of both HVDC links is symmetric, because INELFE interconnector is in multi-link control.

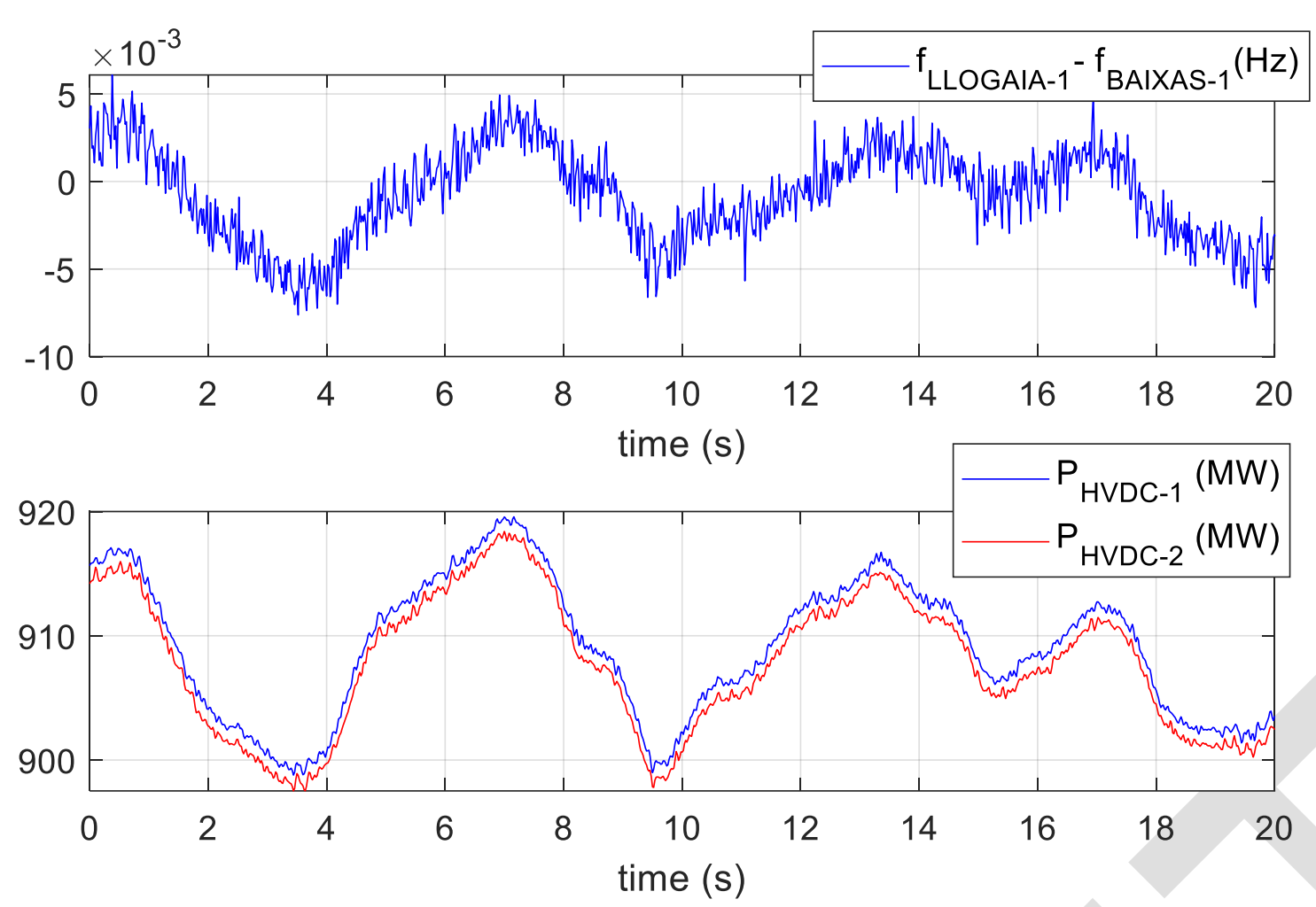

**Figure 6: Test 1-a (CPC). Difference between the frequencies of Llogaia and Baixas substations and the active-power through both HVDC links.**

## Test 1-b: ADC with POD-P

Figure 7 shows the difference between the frequencies of Llogaia and Baixas substations and the active-power through both HVDC links of INELFE-1 (HVDC-1 and HVDC-2) from Spain to France). The behavior of the POD-P controller is the same as the one before, but, in this case, it is superposed to the active-power set point of ADC. In fact, the contribution of ADC is much higher than the contribution of POD-P controller and active-power modulation provided by POD-P controller cannot be distinguished in Figure 7. The test proves the correct behavior of POD-P controller when using ADC mode.

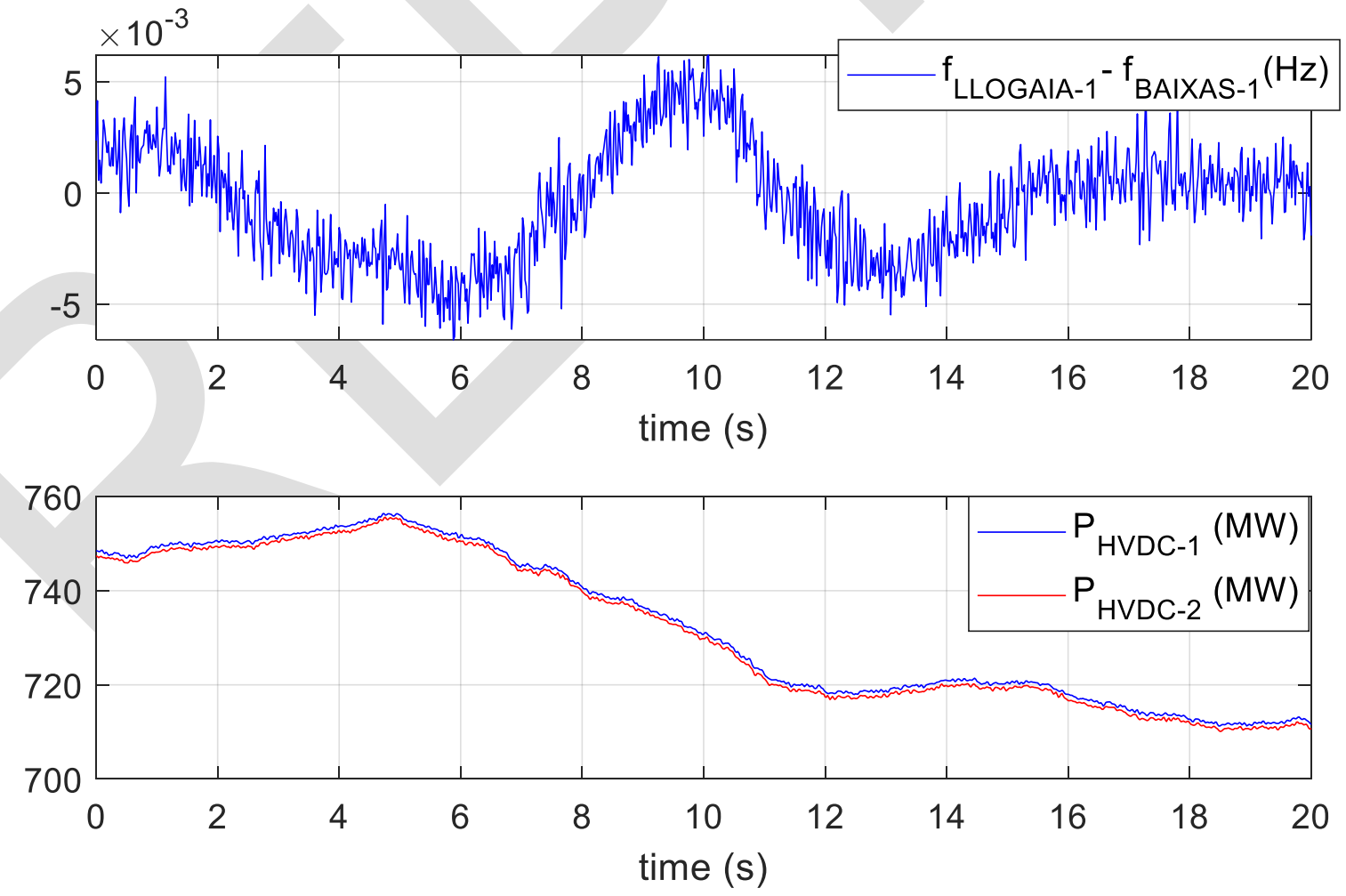

**Figure 7: Test 1-b (ADC). Difference between the frequencies of Llogaia and Baixas substations and the active-power through both HVDC links.**

## Damping ratio of the inter-area mode

The damping ratio of inter-area modes in monitored during the tests, with a Wide Area Monitoring System (WAMS), obtaining real-time data from the PMUs [24]. The critical inter-area mode is the East-Center-West inter-area mode of Continental Europe, with an oscillation frequency of 0.15-0.25 Hz. During the tests, the frequency of this mode obtained from the WAMS is around 0.15 Hz. Figure 8 shows the damping ratio of the inter-area mode obtained from the WAMS (from PMUs located in Spanish power system) during the different tests. At certain points, data were missing and the last available value was assumed. This can be observed in some constant stretches in the damping ratio showed in Figure 8. In addition, the average of the damping ratio of the inter-area mode has been calculated for the time window of each case, which are shown with green color in Figure 8 and they are summarized in Table 4. Results show that the damping ratio of the inter-area mode increases when activating POD-P controller of INELFE HVDC interconnector, for both, CPC and ADC modes. The average damping ratio increases around 10 % when activating the POD-P controller (see Table 4). However, it is important to highlight that the effect of the POD-P controller itself cannot be identified, because the power system conditions have impact on the damping ratio of the inter-area mode.

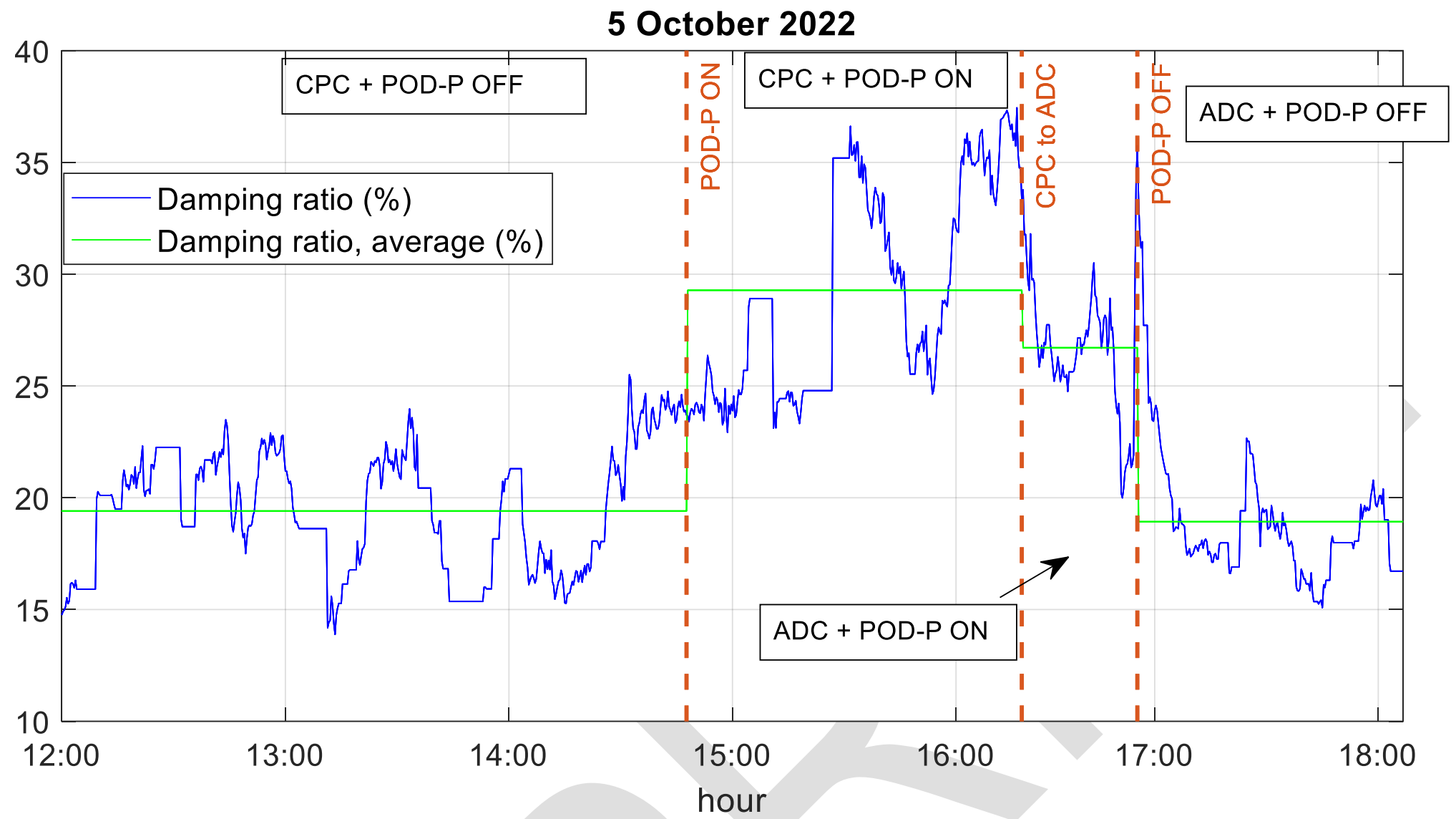

**Figure 8: Damping ratio and frequency of the inter-area mode during the tests carried out between 12 :00 and 18 :00 of 5th October 2022.**

**Table 4 : Average damping ratio of the inter-area mode during the different time windows of the tests (5th October 2022).**

| Test | | Time | Average damping ratio (%) |
|---|---|---|---|
| CPC | POD-P OFF | 12:00h – 14:47h | 19.40 |
| | POD-P ON | 14:47h – 16:17h | 29.28 |
| ADC | POD-P ON | 16:17h – 16:48h | 26.71 |
| | POD-P OFF | 16:48h – 18:00h | 18.93 |

## Test 2: Tests of POD-P control of INELFE-1 under the disconnection of line Vic-Baixas 400 kV (9-Dec-2022)

These tests consist on analyzing the behavior of POD-P in CPC and ADC operating modes, when Line Vic-Baixas 400 kV is disconnected. The objective of this test is to evaluate the improvements that POD-P controller could produce, against a disturbance in the system. Two tests were carried out:

- Test 2-a: CPC Mode. Disconnection of line Vic-Baixas 400 kV.
  - POD-OFF / POD-P ON
- Test 2-b: ADC and POD-P controller. Disconnection of line Vic-Baixas 400 kV.
  - POD-OFF / POD-P ON

The operating conditions of the system before the tests were as follows:

- VSC-HVDC interconnection: HVDC-1 and HVDC-2 in service. Control mode: Multi-link. France is PCONMASTER and active-power flow from Spain to France.
- Line Vic-Baixas 400 kV connected .

Only the results of Test 2-a are presented in this paper, due to the lack of space.

Table 5 shows the active power flow through line Vic-Baixas 400 kV, before the opening the line, for each case tested. Notice that the VSC-HVDC links are exporting power from Spain to France, while AC line Vic-Baixas 400 kV is transferring power from France to Spain. This is due to the fact that total programmed exchange between France and Spain must be maintained during the tests.

**Table 5 : active power flow through line Vic-Baixas 400 kV.**

| Test | | $P_{VIC-BAIXAS}$ (MW) |
|---|---|---|
| Test 2-a: CPC | POD-P OFF | -422.54 |
| | POD-P ON | -328.58 |

**Test 2-a: CPC with POD-P**

Both HVDC links are in CPC mode. Line Vic-Baixas 400 kV is disconnected at t=1 s. Two cases are compared with POD-P OFF and with POD-P ON.

In order to analyze inter-area oscillations in the power system, the frequency recorded in a PMU close to a generator (G1, for short) is analyzed. Figure 9 compares the difference between the frequency generator G1 and the frequency in Baixas 400 kV. The dynamic response with POD-P controller presents a higher damping ratio than the one obtained with POD-P OFF.

Figure 10 shows the difference between the frequencies of Llogaia and Baixas substations and the active-power through HVDC-1 (from Spain to France). Only the results of one link are shown (/HVDC-1), because the behavior of the other link is symmetric. POD-P controller presents a correct behavior: when the frequency of Llogaia is greater (lower) than the frequency of Baixas, the active power through the HVDC links (from Spain to France) increases (decreases).

Results show that the POD-P controller is effective to damp inter-area oscillations. Nevertheless, it should be mentioned that it is difficult to quantify exactly the improvements produced by the POD-P controller, only. This is because power system conditions are different for each test, and all these factors have an impact on the damping ratio inter-area oscillations.

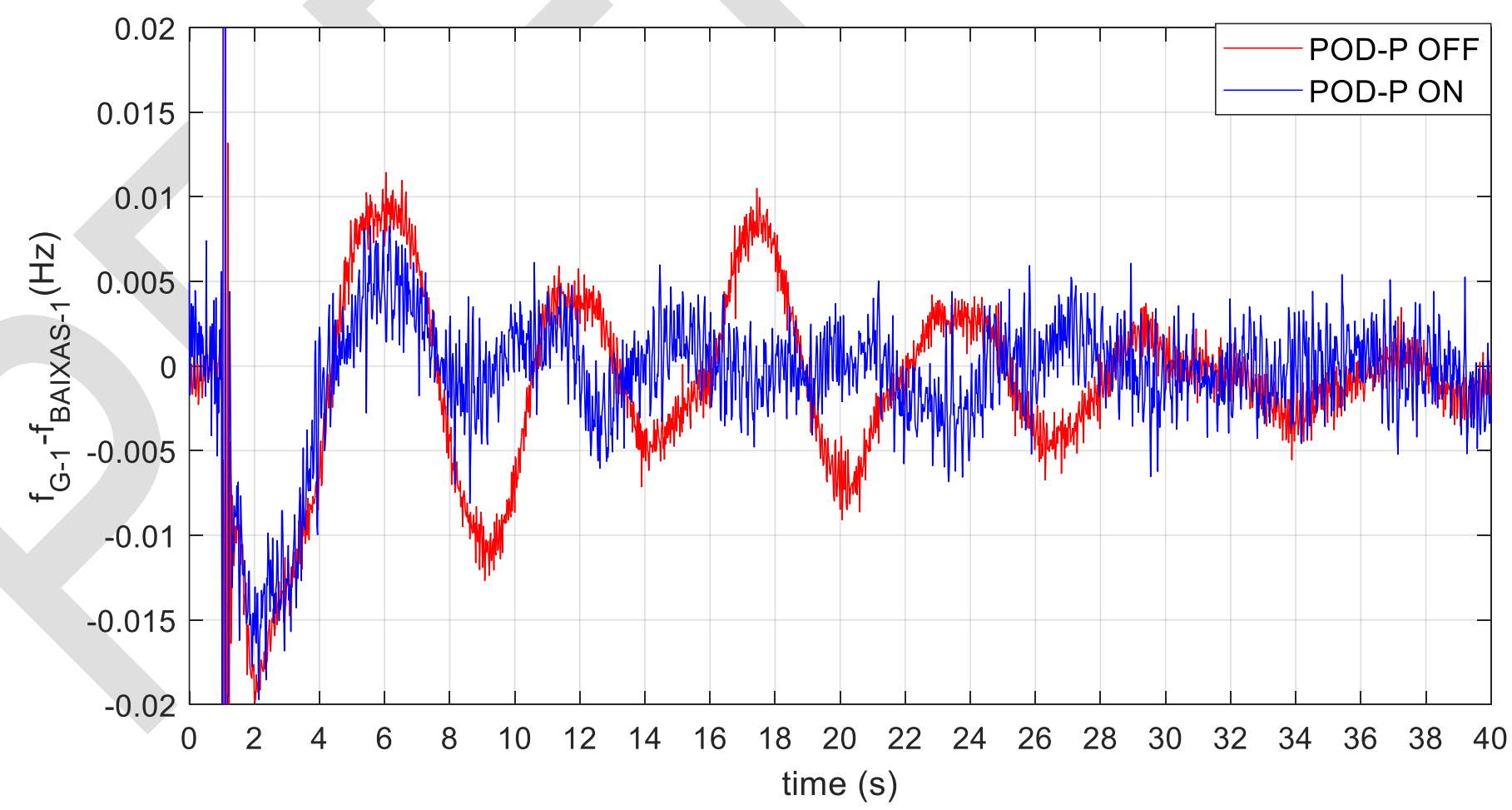

**Figure 9 : Test 2-a (CPC). Difference between the frequency close to a generator in Spain and the frequency at Baixas 400 kV substation.**

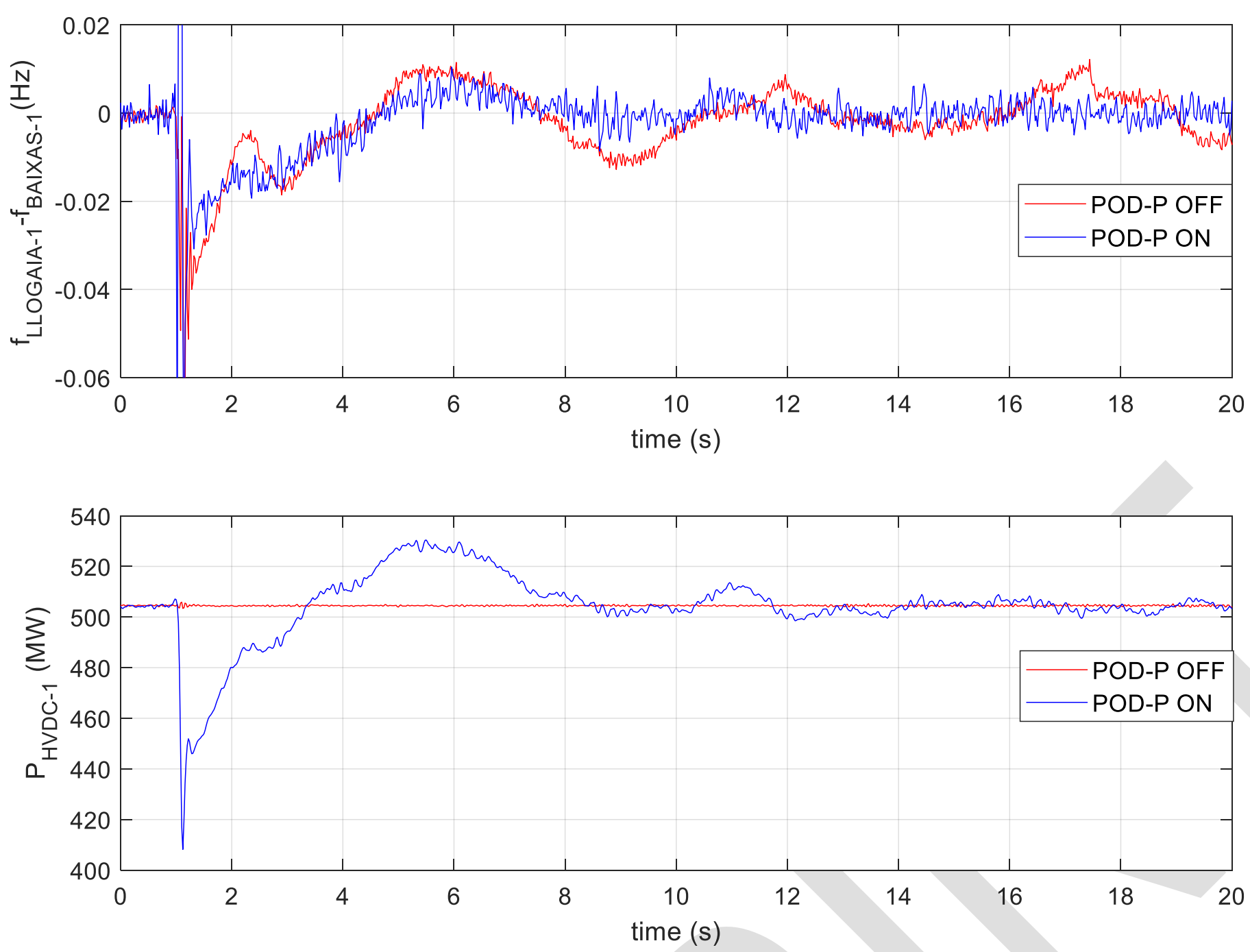

**Figure 10 : Test 2-a (CPC). Difference between the frequencies of Llogaia and Baixas substations and the active-power through HVDC-1.**

## 6 CONCLUSIONS

This paper presented simulation studies, hardware-in-the-loop and field tests on the POD-P controller of INELFE VSC-HVDC France-Spain interconnector, for the validation of the changes made in POD-P controller (increase of the controller gain and to allow the activation of POD-P controller when operating the link in angle difference control (ADC)).

Main conclusions of this paper can be summarized as follows:

- Modifications carried out for POD-P controller of INELFE VSC-HVDC link have been implemented successfully.
- Tests prove correct behavior of POD-P controller in different scenarios and in constant active power control (CPC) and in angle difference control (ADC).
- Results show that POD-P increases the damping ratio of the inter-area mode. Improvements are higher when line Vic-Baixas 400 kV is open.
- POD-P controller with the new changes has been validated and it is being used in normal operation for both, CPC and ADC modes, since 5$^{th}$ January 2023.
- Due to all the factors that provoke changes in the damping ratio of the inter-area mode, it is difficult to quantify exactly the improvements produced by the POD-P controller, only.